# Emergence of ferromagnetic state due to structural disorder in pseudo-binary Ce(Fe$_{0.9}$Co$_{0.1}$)$_2$ compound

A. Musiał, M. Pugaczowa-Michalska, N. Lindner, Z. Śniadecki*

Institute of Molecular Physics, Polish Academy of Sciences
M. Smoluchowskiego 17, 60-179 Poznań, Poland

*corresponding author: Zbigniew Śniadecki, e-mail: sniadecki@ifmpan.poznan.pl

**Abstract**

The changes in magnetic properties of Ce(Fe$_{0.9}$Co$_{0.1}$)$_2$ compound with increasing disorder are discussed in the paper. Homogeneous alloys are known to undergo the phase transition from ferromagnetic to antiferromagnetic state accompanied by the structural distortion of the cubic Laves C15 phase into the rhombohedral one. Various stimuli, like the structural disorder, or applied magnetic field, can force the emergence of ferromagnetism at low temperatures. We initially introduced the structural disorder using rapid quenching. Further changes were made by severe plastic deformation. The presence of a ferromagnetic phase in a low-temperature region is reported here and accompanies the deterioration of a first-order phase transition. We show, based on electronic calculations, that the structural motifs arising from various distortions of the initial MgCu$_2$-type structure, caused by the partial replacement of Fe with Co atoms, are characterized by stable antiferromagnetic order. This neglects simple structural distortions as the source of ferromagnetism. The presence of a strongly defective structure understood as a topologically disordered volume, reduced the fraction transformed from a ferromagnetic to an antiferromagnetic state. Therefore, a strong reduction of isothermal entropy changes was also observed, as it decreased from 1.94 Jkg$^{-1}$K$^{-1}$ and -1.43 Jkg$^{-1}$K$^{-1}$ ($\Delta\mu_0 H$ = 4 T) to 0.30 Jkg$^{-1}$K$^{-1}$ and -0.96 Jkg$^{-1}$K$^{-1}$ for antiferromagnetic-ferromagnetic and ferromagnetic-paramagnetic transition, respectively.

**Introduction**

The REFe$_2$-type alloys (RE = rare-earth element) characterized by MgCu$_2$-type cubic structure attract significant interest due to the variety of interesting properties and possible applications. For example, the Laves phase REFe$_2$ alloys are promising hydrogen storage materials due to their high volumetric hydrogen storage density [1-3]. They are known for their unique magnetocaloric properties in a broad temperature range [4]. Typically, physical phenomena in this system are governed by its specific crystal structure with characteristic tetrahedra of transition metal atoms. According to lanthanides contraction, the lattice parameter should decrease for RE elements with higher atomic numbers. Surprisingly, Ce is the exception here [5] giving rise to unexpected phenomena described in the further part of this work. The lowest lattice parameter in the CeTM$_2$ (TM = Fe, Co, Ni) group of intermetallics was obtained for CeCo$_2$, while the highest one was for CeFe$_2$ [5, 6].

In general, rare-earth atoms provide high magnetic anisotropy due to localized magnetic moments. Depending on the RE element, magnetic moments on the RE site can be ordered parallel to the Fe sublattice, for the light ones, or antiparallel for heavy RE. In the CeFe$_2$ alloy, the magnetic moments of Ce are ordered unexpectedly antiparallel to the magnetic moments of iron, and unusual magnetic behavior is observed. Strong hybridization between Ce 4*f* and Fe 3*d* electrons results in a low ferromagnetic ordering temperature of about 230 K and a magnetic moment equal to 2.4 μ$_B$/f.u. [7]. Additionally, distinct magnetic ground state and behavior can be observed for the REFe$_2$ compounds substituted by Co, Al, Cr, or Ru, leading to the emergence of ferromagnetic (FM) and antiferromagnetic (AFM) phases (separated by a first-order phase transition) or to their coexistence [8-12]. This phase transition from FM to AFM state is accompanied by the structural distortion of the cubic phase (space group *Fd-3m*) into the rhombohedral one (space group *R-3m*) [13, 14]. Similar behavior was observed in the (Tb$_{0.2}$Pr$_{0.8}$)$_{1-x}$Ce$_x$Fe$_{1.93}$ (0 ≤ x ≤ 1.0) for the high Ce content of x > 0.6, where the easy magnetization direction change was accompanied by the structural distortion (from rhombohedral to orthorhombic) [15]. On the other hand, not all substitutions induce magnetostructural transition, as reported for Ag and Au in the CeFeX system [11]. Based on XPS results, Das *et al.* have shown that Ni substitutions in $Ce(Fe_{1-x}Ni_x)_2$, where x = 0; 0.5; 0.8, cause relocalization of Ce $4f$ density of states and results in the reduction of $f$-$d$ hybridization [16]. Mn substitution in the quasibinary $CeFe_{2-x}Mn_x$ compound (where $x > 0.3$), induces a noncollinear magnetic structure in the $3d$ magnetic sublattice, simultaneously causing the decrease of Curie temperature with increasing Mn content and suggesting the presence of heterogeneous exchange interactions [17]. Similarly the increasing content of Si in $Ce(Fe_{1-x}Si_x)_2$ (x = 0-0.07) results in the decrease of Curie temperature. Additionally, it is followed by the appearance of the AFM phase at low temperatures for concentrations of silicon exceeding $x \geq 0.05$. Such high susceptibility of these compounds to doping and external stimuli indicates that the stability of a specific magnetic ordering in the ground state is low and therefore potentially tunable.

Properties of $Ce(Fe_{0.9}Co_{0.1})_2$ are strictly connected with the physical mechanisms and characteristics mentioned above for the whole family of Ce-based Laves phases. Wada *et al.* calculated the electronic specific heat coefficient $\gamma$ for $Ce(Fe_{1-x}Co_x)_2$ in the ferromagnetic and antiferromagnetic state, showing that the mechanism governing variation of the electronic specific heat plays a crucial role in FM-AFM transition, and implying that the thermal properties at low temperatures are dominated by spin fluctuations [18]. Many experimental results reported for $Ce(Fe_{1-x}Co_x)_2$ (also for the compound of interest, $Ce(Fe_{0.9}Co_{0.1})_2$) are consistent with the theoretical predictions of Moriya and Usami [19] on strongly interacting itinerant electron systems and can be attributed to the transfer of the Ce $4f$ electrons to the conduction band [6]. Another example of the high sensitivity of the mentioned phases to the substitution effect is an increase in the Fe-Fe direct exchange interaction and a significant increase in the Curie temperature induced by substituting Ce with nonmagnetic Y. What is also interesting from our perspective is the enhancement of relative cooling power (RCP) resulting from the broadening of the peak observed in the temperature dependence of isothermal entropy change for substituted alloys [20]. As already mentioned, even small variations in chemical composition or the presence of chemical disorder strongly influence the resulting electronic or magnetic properties of Fe-based Laves phases.

To make a further step, the main aim of the present work is to show the changes in the magnetic properties with increasing topological disorder. The disorder was initially introduced by rapid quenching, while further, more significant changes were caused by severe plastic deformation. We believe the structural disorder can substantially enhance ferromagnetic ordering as reported for another itinerant electron system [21] and significantly impact the magnetocaloric properties of $Ce(Fe_{0.9}Co_{0.1})_2$ compound. The *ab initio* calculations presented are intended to prove whether the structural units that can be modeled (*i.e.* a series of supercell models) are responsible for the appearance of ferromagnetism in this system.

**Experiment and calculations**

The ingot of $Ce(Fe_{0.9}Co_{0.1})_2$ was prepared by arc-melting of high-purity elements (3N or more) in an argon atmosphere. It was remelted several times to ensure homogeneity. The weight loss was determined to be less than 0.1 wt.% and was connected with the evaporation of Ce. The sample was then rapidly quenched by melt-spinning technique (wheel surface velocity was equal to 30 m/s). The disk-shaped samples of 10 mm in diameter were processed by high-pressure torsion (HPT) using a computer-controlled HPT machine (W. Klement, Lang, Austria) and between two flat anvils under quasi-hydrostatic pressure of 6 GPa at room temperature. The sample was processed for one revolution with a rotational speed of 0.2 revolutions per minute. Structural data were obtained by x-ray diffraction (XRD) with the use of a PANalytical

X'Pert diffractometer in Bragg-Brentano geometry with CuKα radiation ($\lambda$ = 1.5406 Å) over the 2$\Theta$ range of 20°–80°, at 0.017° step size. Temperature dependences of magnetization were measured on the vibrating sample magnetometer (VSM) option of Quantum Design PPMS from 2 to 320 K in the applied magnetic field of $\mu_0H$ = 0.5 T. Magnetization isotherms were measured between 2 and 300 K in magnetic fields up to 6 T. Magnetization curves were further used to calculate temperature dependence of isothermal entropy change $\Delta S(T)$.

Calculations were performed by an *ab initio* plane wave pseudopotential code (VASP) [22, 23]. Pseudopotentials generated using the projector augmented wave (PAW) method [24] were used throughout the calculation together with the generalized gradient approximation (GGA) in the Perdew, Burke, and Ernzerhof (PBE) [25] parametrization. The cutoff energy for plane waves was 520 eV throughout the calculations, with the convergence criteria being $10^{-6}$ eV for electronic iterations. The supercell model was used to simulate disorder as introduced to crystal structures of CeFe$_2$ by substituting Co on Fe sites. We used a supercell with eight formula units (f.u.) of the parent CeFe$_2$ system with the cubic structure *Fd-3m* (space group nr. 227). A 4x4x4 k-points grid was used in the Monkhorst-Pack scheme [26] in supercell calculations.

**Results and discussion**

Figure 1 shows the x-ray diffraction patterns measured at 300 K for the as-quenched and HPT-treated samples of Ce(Fe$_{0.9}$Co$_{0.1}$)$_2$. Both can be identified as referring to the same structure with high symmetry. Going into the details, the short vertical indicators on the bottom bar in Fig. 1. correspond to the calculated peak positions for the MgCu$_2$-type cubic structure (C15 Laves phase). Therefore, we can confirm that the sample is almost a single phase after the rapid quenching process. There is one distinct Ce-oxide peak at $2\theta$ = 30º, reported earlier *e.g.* by Fukuda *et al.* [9]. Ce is easily oxidizing and the presence of oxide is expected especially since the sample for the diffraction experiment was ground into a fine powder. Ce-oxide was not detected in the high-pressure torsion-treated sample, because a solid piece was cut for an x-ray diffraction experiment and the surface/bulk ratio was minimized. The lattice parameter determined for the as-quenched sample is equal to $a = 7.313 \pm 0.001$ Å and is higher than the value reported by Koyama *et al.* ($a = 7.293$ Å) [13]. This may be due to lattice expansion caused by a chemical disorder or uniform lattice strain. The large broadening of diffraction peaks leads to difficulties in the Rietveld analysis for the HPT-processed sample and should be treated as tentative. Nevertheless, the presence of strongly refined grains and structural disorder is evident. Such line broadening was expected, as we reported this effect for the other C15 Laves phase [21] and then attributed it to the presence of small nanometric grains, microstrains, and chemical/topological disorder within the crystalline structure.

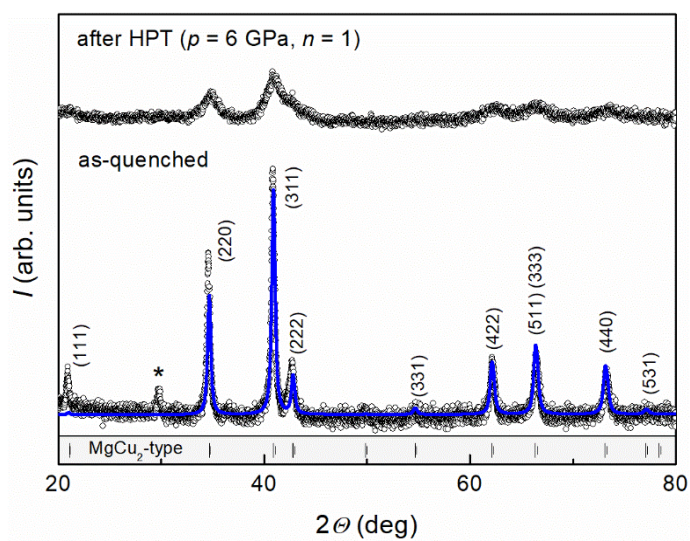

Figure 1. X-ray diffraction patterns measured at 300 K for Ce(Fe$_{0.9}$Co$_{0.1}$)$_2$ in an as-quenched state and after plastic deformation (HPT—high-pressure torsion, $p$ – pressure, $n$ – number of revolutions).

Assuming that the MgCu$_2$-type phase is also dominant in the plastically deformed sample, we estimated the lattice parameter to be equal to $7.300 \pm 0.001$ Å. One should bear in mind that this value was determined not considering the presence of *e.g.* uniform microstrains, which are highly expected in this case and could strongly shift the positions of the diffraction peaks. Nevertheless, the strongly reduced value of the lattice parameter for the HPT-treated sample (compared to the as-quenched one) can also be attributed to other effects, *e.g.* the chemical ordering of the C15 Laves phase. It has already been confirmed by neutron and x-ray diffraction that the phase transition from ferromagnetic to antiferromagnetic state accompanies the structural distortion of the cubic phase (space group *Fd-3m*) into the rhombohedral one (space group *R-3m*) [13, 14]. Kennedy *et al.* reported [14] that the rhombohedral distortion in Ce(Fe$_{0.8}$Co$_{0.2}$)$_2$ is characterized by the inter-axis angle $\alpha = 90.20° \pm 0.01°$ at 18 K. Similar value, $\alpha = 90.32°$ at 50 K, was reported by Koyama *et al.* for Ce(Fe$_{0.9}$Co$_{0.1}$)$_2$ [13]. Assuming the high similarity of cubic and rhombohedral patterns [13] and the presence of structural disorder (peak broadening) in both analyzed samples, we ruled out x-ray diffraction as decisive in this case. In short, it is impossible for the deformed sample, to distinguish the presence of cubic or other distorted structures *e.g.* rhombohedral one. Therefore, additional first principles calculations were conducted by us and the results are further presented and discussed along with the analysis of magnetic properties.

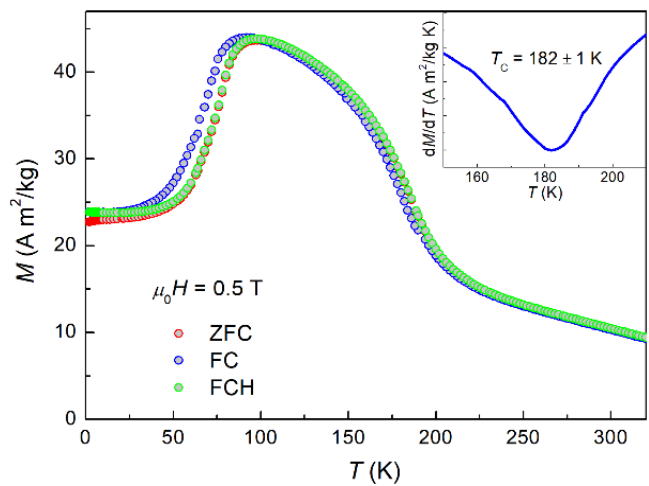

Figure 2. Temperature dependence of magnetization of Ce(Fe$_{0.9}$Co$_{0.1}$)$_2$ in an as-quenched state measured in 0.5 T (ZFC – zero field cooled, FC – field cooling, FCH – field cooled heating).

Temperature dependence of the magnetization of Ce(Fe$_{0.9}$Co$_{0.1}$)$_2$ in an as-quenched state is shown in Fig. 2. Magnetization curves were measured in various modes *i.e.* after cooling without external magnetic field (ZFC), during cooling in an external magnetic field of 0.5 T (FC) and heating after the sample was firstly cooled in 0.5 T (FCH). While the chosen magnetic field is relatively strong, and not ideal for determining transition temperatures, it was selected to enable the most comprehensive comparison with literature. Starting from the high-temperature range, the sample shows a second-order transition from the paramagnetic (PM) to ferromagnetic (FM) state at Curie temperature $T_C$ = 182 ± 1 K. This value was estimated based on the temperature dependence of the derivative of the FC magnetization curve (inset in Fig. 2). It should be noted here that the *M*(*T*) curve changes around $T_C$ in a rather typical way, apart from high magnetic residue above 200 K. Such high temperature behavior would be expected for stronger magnetic fields, but for 0.1 T magnetization has been shown to decrease rapidly almost to zero in polycrystalline homogenized Ce(Fe$_{0.9}$Co$_{0.1}$)$_2$ [13]. Therefore, we believe the observed effect is connected with a ferromagnetic phase that was not detectable in x-ray diffraction *e.g.* atomic clusters with higher $T_C$ than the parent phase. Such ferromagnetically ordered units are likely in highly disordered alloys where various Fe nearest neighbor surroundings are expected, with the bcc Fe structure as a compositional extremum. Finally, for Ce(Fe$_{0.9}$Co$_{0.1}$)$_2$ one should expect to observe similar behavior as reported for parent CeFe$_2$, along with possible ferromagnetic ordering of surface Ce moments (especially with increasing grain refinement) or in general extreme sensitivity of the Ce 4*f* states to the external perturbations [27]. In the end, the character and temperature of the observed FM-PM transition in Ce(Fe$_{0.9}$Co$_{0.1}$)$_2$ are consistent with the literature [10, 11, 13].

In CeFe$_2$ the FM phase has a cubic structure (space group *Fd*-3*m*). The antiferromagnetic one (AFM) is linked with a rhombohedral crystalline structure (space group *R*-3*m*) and is observed at low temperatures for the small substitution of Fe with Co (or other metals, *e.g.* Ru, Re, or Al) [8, 28, 29]. This is the case here, where at low temperatures an antiferromagnetically ordered state emerges. The FM-AFM transition is visible as the abrupt decrease of magnetization below ~90 K (see. Fig. 2), as already reported by Koyama *et al.* [13]. This temperature is also consistent with the onset of nucleation of the AFM phase upon cooling in Ce(Fe$_{0.95}$Ru$_{0.05}$)$_2$ [8]. Thermal hysteresis visible in Fig. 2 (about 10 K in 0.5 T) is an indication of a first-order phase transition. The same value has been reported by Rastogi and Murani for a 20% substitution of Co [30]. Moreover, when the temperature of the onset of the AFM phase nucleation on cooling is higher than the onset of the FM phase nucleation upon heating, which is the case here, the first-order phase transition is reported to be influenced by disorder [8]. We may assume here that the coexistence of FM and AFM phases takes place over a wide range of temperatures defined at bifurcation points of ZFC and FC magnetization curves. Even though it was not possible to confirm it here by x-ray diffraction (due to high structural disorder and peak broadening), such evidence has been given by Koyama *et al.* [13]. Moreover, Roy *et al.* confirmed the temporal growth of the clusters inside the phase-coexistence regime indicating the phases' metastability and sensitivity to energy fluctuations, as another evidence of disorder-influenced first-order phase transition [31, 32]. All these features are common for the systems with quenched disorder and impact the competition between ordered states separated by a first-order transition.

Leaving aside for a moment the observations consistent with the literature, attention should also be drawn to the differences. The magnetization measured in ZFC mode does not reach values close to zero, as expected in such antiferromagnets [13]. This could be the effect of an external magnetic field, which deteriorates the AFM state. In the literature, even though the antiferromagnetic phase was not observed in an external magnetic field of 9 T [13], the ZFC curve reached values close to zero at low temperatures for 5 T already, in contradiction to the $M(T)$ ZFC curve measured by us in 0.5 T. Therefore, we have to link this effect with something else, namely the presence of a ferromagnetic phase down to the lowest temperatures. It was reported that the ferromagnetic state in the C15 Laves phase becomes unstable and the antiferromagnetic state becomes more stable by the lattice contraction [33]. As the lattice parameter in our case is higher than expected for a stable compound, we should expect to observe the stabilization of the ferromagnetic state. Such behavior was reported for the as-quenched sample. On the contrary, the decrease of lattice parameter in the plastically deformed sample (regardless of the origin of this shift) is not consistent with the further stabilization of the ferromagnetic state at low temperatures, as observed in Fig. 3. Magnetization curves measured in FC and FCH mode just slightly deviate from ZFC curve and the value at 2 K (lower measurement limit) is comparable to the maximum above AFM-FM phase transition. This indicates that the vast majority of the sample's volume fraction is ferromagnetic. Therefore, the presence of quenched-in disorder in the form of ferromagnetic clusters/phase, defined as being not entirely structurally consistent with the Laves C15 phase, seems to be the more rational cause of the presence of a ferromagnetic state. The mentioned structural changes that affect the AFM-FM transition in such a significant way are not reflected in the FM-PM transition, as the Curie temperature equal to $186 \pm 1$ K is comparable to $T_C$ determined for the as-quenched sample. The only noticeable difference, naturally connected with a structural disorder, is the broadening of the phase transition range (evident when comparing the width of the d$M$/d$T$($T$) minima in the insets of Figs. 2 and 3).

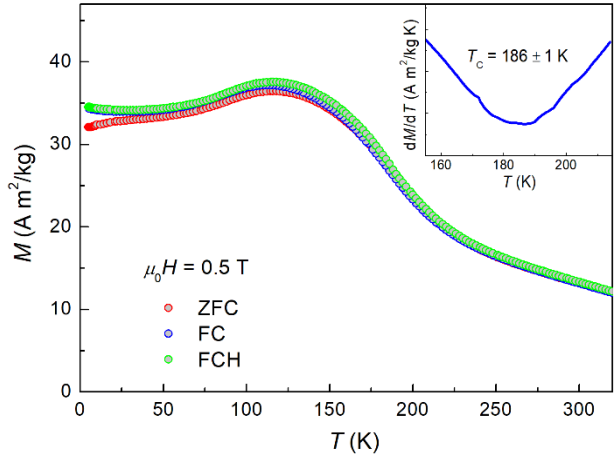

Fig. 3. Temperature dependence of magnetization of Ce(Fe$_{0.9}$Co$_{0.1}$)$_2$ after high-pressure torsion measured in 0.5 T (ZFC – zero field cooled, FC – field cooling, FCH – field cooled heating).

Magnetization isotherms measured for the as-quenched sample (Fig. 4) follow the previous results and conclusions drawn. The $M(\mu_0H)$ curves measured below the AFM-FM transition temperature, or to be more precise below the onset temperature of the AFM phase nucleation, are shown on the left panel in Fig. 4. The results differ from those previously presented in the literature [8, 13], but are in line with the curves measured for the lower content of Co, namely Ce(Fe$_{0.95}$Co$_{0.05}$)$_2$ [9].

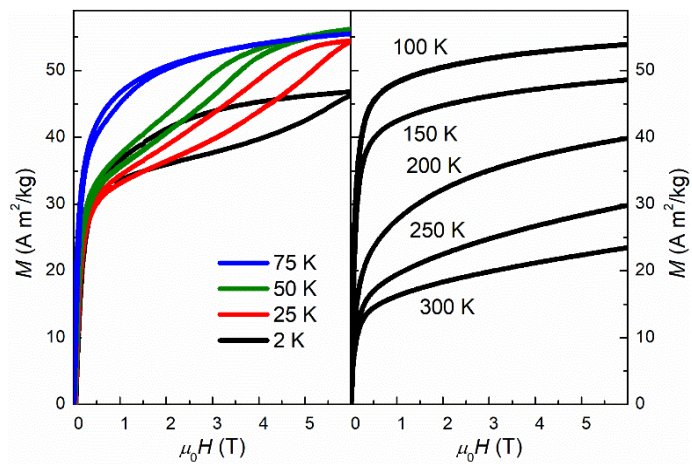

Fig. 4. Magnetization isotherms of Ce(Fe$_{0.9}$Co$_{0.1}$)$_2$ in an as-quenched state measured below (left panel) and above (right panel) the AFM-FM transition temperature.

An abrupt increase of magnetization is observed in low magnetic fields, resembling the isotherms of ferromagnetic materials, and indeed such behavior was already reported to originate from the ferromagnetic component [9]. Metamagnetic phase transition is observed in higher magnetic fields, where the antiferromagnetic state (some volume fraction of antiferromagnetically ordered phase which is present at low temperatures) transforms into an induced ferromagnetic state. The antiferromagnetically ordered phase is rather unstable as the transition starts below 1 T already. Moreover, the fraction of the ferromagnetic phase in the low magnetic field region is much larger than reported for $Ce(Fe_{0.95}Co_{0.05})_2$ [9]. In the latter, the presence of a ferromagnetic fraction was justified by the presence of a canted spin state that may be induced as a result of the competition between the antiferromagnetic correlation arising from Ce $4f$ – Fe $3d$ hybridization and the Fe $3d$ – Fe $3d$ ferromagnetic exchange interaction. If this were to be an explanation of the presence of the ferromagnetic fraction in our alloy, then the Co content in the C15 Laves phase should be lower than assumed, and consequently, this should be reflected in the values of the transition temperatures. However, this is not observed in our system and the values are practically identical to those for the 10% substitution of Co for Fe, as described before. The presence of a canted spin state and the competition of the FM and AFM states (arising through a competition between the ferromagnetic Fe $3d$–Fe $3d$ interaction, and the antiferromagnetic interaction originated from the hybridization among Ce $4f$–Ce $5d$–Fe $3d$ [34]), is not excluded, but the dominating effect in our case must arise due to the presence of structural units of a different type than the mentioned compositional modification of the cubic phase (depletion of Co content). In conclusion, the ferromagnetic state is not an intrinsic effect of the primary cubic structure and is connected with the presence of structural units/defects introduced by non-equilibrium synthesis methods. Some signatures of nucleation of nanosized ferromagnetic domains in the vicinity of structural defects in the $MgCu_2$-type phase (caused by neutron irradiation) were found in Re-containing samples [35, 36]. It was reported that these domains remained frozen (in FM state) across the transition to the AFM state, but the magnetization related to their presence reached rather meager values (less than 0.1 emu/g). That is several orders of magnitude lower than the *M* values reported here (Figs. 2 and 3). Therefore, we opt for a highly defective structure and the presence of strong distortions, rather than just the simple creation of defects within the C15 Laves phase structure.

Above 90 K, the magnetization isotherms (right panel in Fig. 4) show a typical ferromagnetic behavior. All the magnetization isotherms measured for the plastically deformed sample (Fig. 5) are characteristic of ferromagnets. The antiferromagnetically coupled phase almost completely deteriorated upon further growth of structural disorder compared to the as-quenched sample.

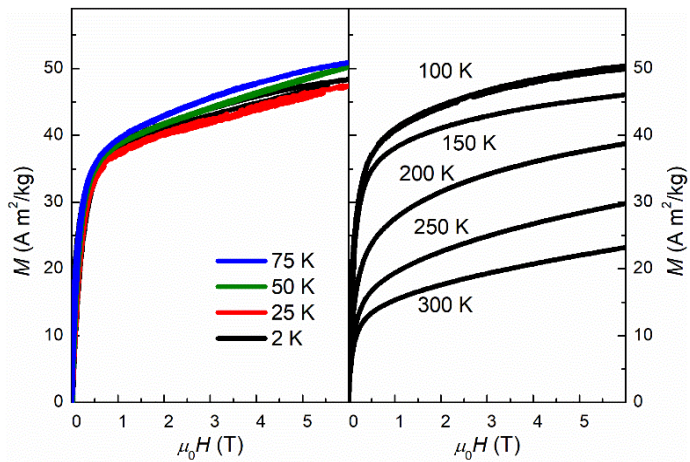

Fig. 5. Magnetization isotherms of plastically deformed (HPT-treated) Ce(Fe$_{0.9}$Co$_{0.1}$)$_2$ measured below (left panel) and above (right panel) the AFM-FM transition temperature.

Much effort has been put into correctly describing the role of Ce in these systems, especially to predict mechanisms of the formation of the magnetic ground state and the role of Ce, but also Fe, and substitutional Co atoms in this process. The antiparallel ordering of the Ce and Fe magnetic moments in CeFe$_2$ is unusual for the LREFe$_2$ (LRE - light rare earth element) series. The existence of a magnetic moment on Ce was confirmed many times experimentally and theoretically and was reported to be antiferromagnetically coupled to Fe with additional quenching of the orbital 4$f$ magnetic moment [6]. Many experiments have confirmed the antiferromagnetic coupling and allowed the determination of the magnetic moment on cerium, ranging from -0.57 to -0.14 $\mu_B$ [37-41]. Nevertheless, because of the relatively small magnetic moment of the Ce atom, this compound is usually treated in the literature as a ferromagnet [9, 17]. The importance of the hybridization effect between Ce 4$f$, Ce 5$d$, and Fe 3$d$ electrons on the formation of a ground state in CeFe$_2$ was already mentioned by us and indicates that the origin of magnetism in such a system is not strictly connected with the Ce-sublattice, but is governed by the instabilities in the Fe-sublattice [9].

Bearing in mind such instability and possible sensitivity to structural distortions, and trying to find the origin of the emergence and predominance of the ferromagnetic state below AFM-FM transition temperature, we performed *ab initio* calculations. As suggested before, due to the strong affinity of cubic and rhombohedral x-ray diffraction patterns [13] and significant peak broadening in both analyzed samples (Fig. 1), we assumed that x-ray diffraction analysis of crystal structure is not conclusive and adequate in the case of such highly disordered alloys. From the experimental part, we deduced that the origin of the ferromagnetic state at low temperatures is connected with the presence of structural disorder. Most importantly, structural disorder is understood not as small lattice distortions of C15 Laves phase, but entirely new areas with altered structure/topology such as for example disordered grain boundaries or amorphous phase. Nevertheless, let us check if we can also find some elementary structural motifs in the form of cubic phase distortions with a ferromagnetic ground state. The substitutional disorder was introduced using the supercell method within the structure containing eight formula units (f.u.). We assumed that the substitutional disorder in Ce(Fe$_{1-x}$Co$_x$)$_2$ can be modeled as a mixture of several supercells. The composition of such supercells, namely Ce(Fe$_{0.875}$Co$_{0.125}$)$_2$, is close to the stoichiometry of experimentally analyzed samples.

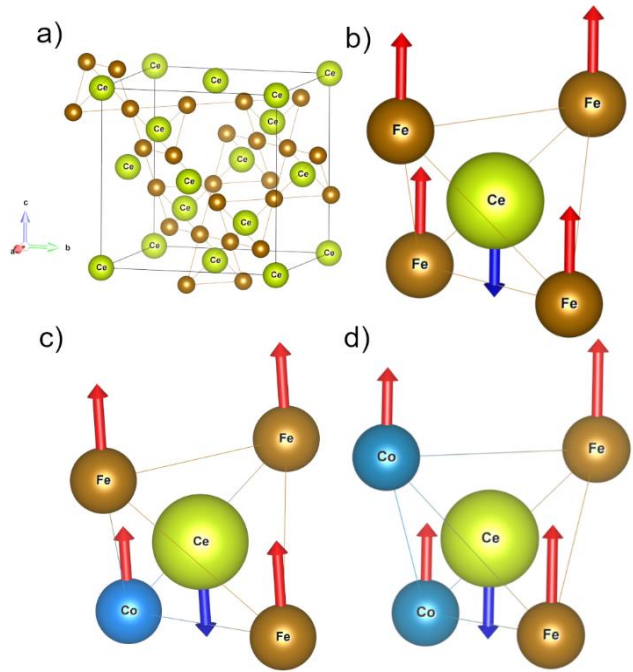

Fig. 6. Schematic picture of the CeFe$_2$ unit cell (a) and three types of various nearest neighborhoods of Ce that can be found in the Ce(Fe$_{0.875}$Co$_{0.125}$)$_2$ supercell for small Co substitutions (b, c, d).

Our supercells take into account various systems with two Co atoms instead of Fe ones. First, two Co atoms replaced Fe atoms in the same CeFe$_4$ tetrahedron, then were located in the nearest neighboring tetrahedra, and in subsequent stages - in all further apart octahedra present in the supercell. Tetrahedra visible as distinct structural units in the schematic picture of the initial cell (Fig. 6a), can be treated as the smallest structural motif in further consideration and presentation of the Co substitution effect. Thus, three types of various nearest neighborhoods of Ce can be found in the supercell (shown in Figs. 6b, c, and d) reflecting a relatively small substitution of Fe by Co atoms. Eventually, there are six possible arrangements of Ce(Fe$_{0.875}$Co$_{0.125}$)$_2$ supercell. For these six supercell variants, our electronic structure calculations, which assumed collinear ferromagnetic order in the magnetic sublattices, converged to the situation with the alignment of the magnetic moments on Ce, Fe, and Co atoms shown in Figures 6b, c, and d. Thus, this may indicate that it is difficult to stabilize the FM state of the parent phase by replacing Fe with Co. Nevertheless, we considered the antiferromagnetic alignment of the magnetic moments on Ce and transition metal atoms (i.e. Fe and Co) in six supercells separately in our calculations. Most importantly, these antiferromagnetic alignments of magnetic moments, hereinafter referred to as AFM state, are more stable in each case. Therefore, we will further focus on the results obtained for AFM. The energy difference between various structures (or supercell arrangements) is so insignificant (up to 3 meV/f.u.) that there is a high probability of simultaneous formation of all of them in the real system. The initial symmetry (*Fm-3m*) of the crystal changes depending on the transition metal site of Co atoms substitution. With Fe by Co substitution, the analyzed system transforms from cubic *Fm-3m* symmetry into the orthorhombic *Cmm*2, orthorhombic *C*222$_1$, or monoclinic *P*2/*m* one. The energy diagram in Fig. 7 shows how the energy of a supercell changes depending on the distance between Co atoms in such a cell. The supercell, in which Co atoms are spaced about 5.71 Å apart has an overall energy lower than other supercells. We assume that the energy of this cell is zero and according to this assumption, the energy of the remaining supercells was calculated. Calculated lattice parameters and angles are reported in Table 1 (Appendix) for AFM variants.

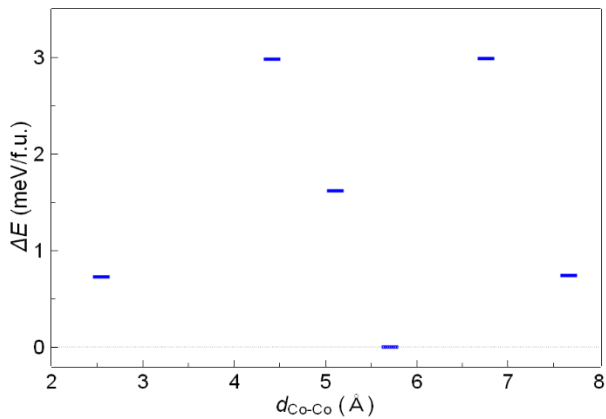

Fig. 7. The difference in total energy between Ce(Fe$_{0.875}$Co$_{0.125}$)$_2$ supercells as a function of distance between Co atoms within them. The energy differences (*ΔE*) were converted to f.u.

The density of states plots shown in Fig. 8 illustrate the total and partial electronic states contributions of Ce(Fe$_{0.875}$Co$_{0.125}$)$_2$ for supercell with $d_{Co-Co}$ = 5.71 Å.

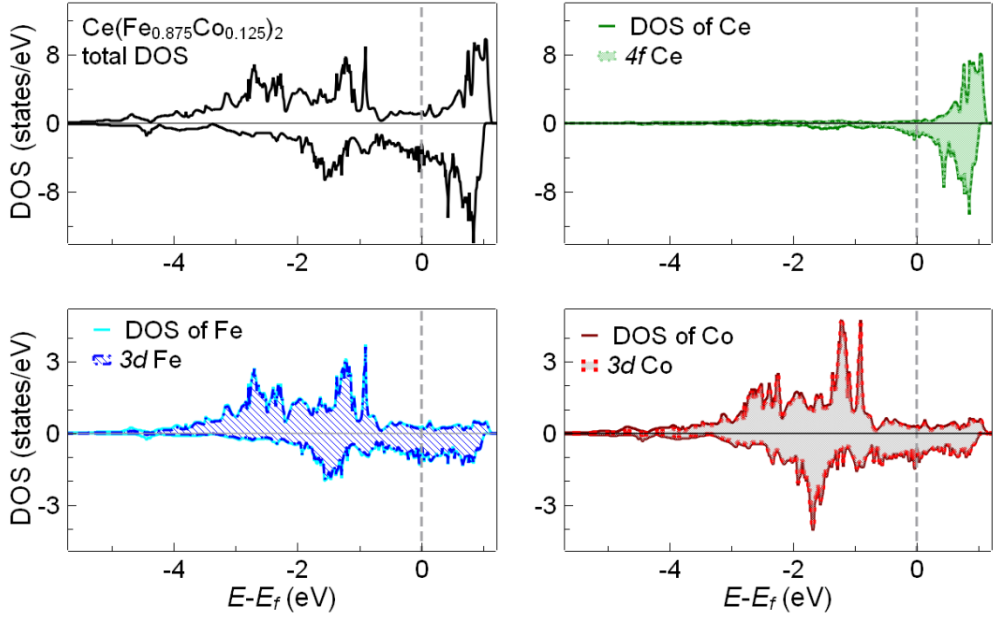

Fig. 8. The total densities of states (DOS) of Ce(Fe$_{0.875}$Co$_{0.125}$)$_2$ and atomic DOS and site-spin-projected electron-state densities: 4$f$-Ce, 3$d$-Fe, 3$d$-Co. The grey dashed line is the Fermi energy.

The total DOS (Fig. 8) is a sum of the individual atomic contributions, from which the dominant role is played by Fe 3$d$ and Co 3$d$ states as well as Ce 4$f$ contribution below Fermi energy ($E_F$). In the region from -1 eV up to the $E_F$ and even higher, these electronic states overlap well, indicating the effect of hybridization of 3$d$ states of Fe and Co with 4$f$ of Ce. Therefore, both in the parent CeFe$_2$ [6] and in Ce(Fe,Co)$_2$, the 4$f$ of Ce have an itinerant character. Konishi and co-authors [42] previously showed that the combined analyses of the experimental core-level and valence-band photoemission, inverse photoemission as well as soft x-ray absorption techniques results, together with theoretical results from band structure calculation within the full-potential linear muffin-tin orbital method (FP-LMTO), indicate a very strong hybridization of the Ce 4$f$ states with the delocalized band states, mainly Fe 3$d$.

Most importantly, the AFM state is more stable in each case. When we consider the smallest structural units, shown in Fig. 6b, c, and d, the values of magnetic moments on Fe, Co, and Ce are very similar in each case. In the Co-less case, the value of the magnetic moment on the Ce atom equals 0.74 $\mu_B$ and is directed opposite to the Fe one (1.73 – 1.75 $\mu_B$, depending on the occupation site). The values of magnetic moment determined for atomic configurations shown in Fig. 6c and d are equal $\mu_{Ce}$ = -0.73 $\mu_B$ and $\mu_{Fe}$ = 1.73 - 1.75 $\mu_B$ in both cases. In turn, $\mu_{Co}$ = 0.66 $\mu_B$ and 0.61 $\mu_B$ depending on the crystallographic site occupied by Co. The most important message is that the FM state is not stable in the case of simple Co-substitution despite changed symmetry resulting from the substitution process and taken by us into consideration.

In conclusion, the literature reports on rhombohedral distortion and the results of our *ab initio* calculations indicate that the AFM state seems stable irrespective of the structural distortion of the MgCu$_2$-type structure. The mentioned observations suggest that this is not the cause of emerging ferromagnetism in our case and that most probably the presence of a strongly defective structure, understood as a volume with topological disorder, is responsible for the unique behavior at low temperatures. In this sense, there is a high similarity between Ce(Fe,Co)$_2$ and another itinerant electron system YCo$_2$, which was previously analyzed by us [21, 43]. In this case, we observe stabilization of the FM phase at the expense of the AFM one. In the latter, we proved that quenched-in disorder and the effects accompanying plastic deformation are responsible for stabilizing the FM phase in the Pauli paramagnet. Both systems are on the verge of being ferromagnetic and even small instabilities (*e.g.* structural disorder, or applied magnetic field) force the stabilization of the FM phase.

One of the aims of this research was also to determine the influence of structural disorder on the magnetocaloric properties of Ce(Fe$_{0.9}$Co$_{0.1}$)$_2$. We wanted to explore the possibility of broadening the temperature range of both observed phase transitions (to increase the full width at half maximum of the $\Delta S$(T)) and analyze its influence on relative cooling power (as a performance metric to rank magnetocaloric materials, indicating heat possible to be extracted in a thermodynamic cycle) [44] and isothermal entropy changes. It has been shown for some Laves phases already [4, 20, 45-47] that the transition temperatures are sensitive to composition changes and that the magnetocaloric effect can be tuned rather easily. Besides, maximum magnetic entropy changes for some of the Fe-based phases of similar composition reached significant values *e.g.* 17 Jkg$^{-1}$K$^{-1}$ for Ce$_{0.9}$Dy$_{0.1}$Fe$_{1.9}$ ($\Delta\mu_0 H = 5$ T) [46].

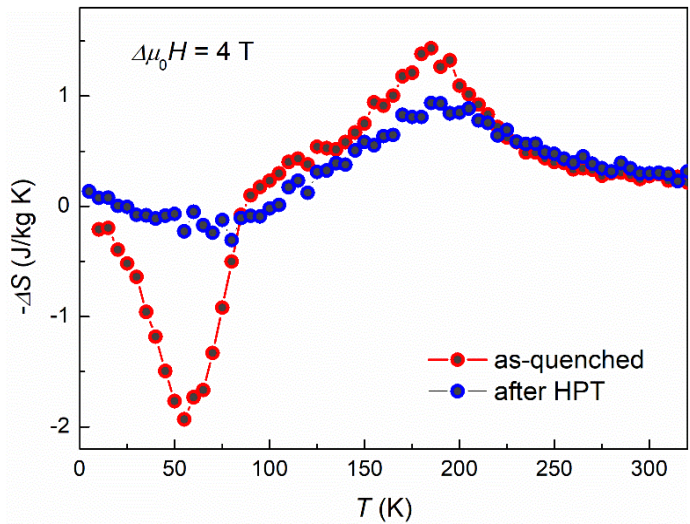

Fig. 9. Temperature dependence of isothermal entropy change ($\Delta\mu_0 H = 4$ T) measured for Ce(Fe$_{0.9}$Co$_{0.1}$)$_2$ in an as-quenched state and after HPT (1 revolution of stacked as-quenched ribbons in 6 GPa).

Figure 9 shows the temperature dependence of isothermal entropy change for the as-quenched and plastically deformed Ce(Fe$_{0.9}$Co$_{0.1}$)$_2$ alloys. Maximum values of $\Delta S$ were reached for the as-quenched sample in the vicinity of both observed phase transitions and are equal to 1.94 Jkg$^{-1}$K$^{-1}$ (AFM-FM) and -1.43 Jkg$^{-1}$K$^{-1}$ (FM-PM) for $\Delta\mu_0 H = 4$ T. Positive sign of isothermal entropy change is characteristic for AFM-FM and negative for FM-PM transition. The $\Delta S$ values decreased rapidly for the plastically deformed sample, to about 0.30 Jkg$^{-1}$K$^{-1}$ and -0.96 Jkg$^{-1}$K$^{-1}$ for AFM-FM and FM-PM transitions, respectively. A structural disorder, introduced by the high-pressure torsion, drastically limited the volume fraction that transformed from a ferromagnetic to an antiferromagnetically coupled phase. This is the cause of the strong reduction of $\Delta S$, by about 85%, compared to the as-quenched sample. Then, the reduction in isothermal entropy changes in the vicinity of the AFM-FM transition temperature is caused by the incomplete structural transformation associated with the presence of the ferromagnetic phase at the lowest temperatures. This reduction is more significant for first-order magnetostructural phase transitions, where two contributions play a role [48]. The contribution connected with magnetic phase change is coupled with the second one, arising from the difference in entropy of two crystallographic structures (rhombohedral and cubic phase in this case).

Structural or magnetic inhomogeneities usually harm magnetocaloric performance, as reported for similar systems showing magnetostructural transition [35]. Isothermal entropy change connected with the FM-PM transition is reduced by about 33 % after the plastic deformation of the as-quenched sample. Despite our plans aiming to improve relative cooling power, the decrease in the value of entropy changes is not compensated by a large broadening of the peak. The relative cooling power values were determined for the as-quenched sample only ($RCP_{AFM-FM}$ = 76.4 J kg$^{-1}$, $RCP_{FM-PM}$ = 104.1 J kg$^{-1}$) and are comparable to those reported for similar alloys [20]. Refrigerant capacity values were also calculated to give more comprehensive view and are slightly lower than RCP with $RC_{AFM-FM}$ = 63.0 J kg$^{-1}$ and $RC_{FM-PM}$ = 82.1 J kg$^{-1}$. In general, introducing structural disorder harms both the magnetocaloric effects associated with the AFM-FM and FM-PM transformations. $\Delta S$ for many similar but homogenized systems reached significantly higher values. For Ce(Fe$_{0.95}$Si$_{0.05}$)$_2$, $\Delta S$ is close to 3 Jkg$^{-1}$K$^{-1}$ and 2 Jkg$^{-1}$K$^{-1}$ (in $\Delta\mu_0 H$ = 4 T) for AFM-FM and FM-PM transitions, respectively [49]. Respective $\Delta S$ values of about 7.4 Jkg$^{-1}$K$^{-1}$ and -3 Jkg$^{-1}$K$^{-1}$ (in $\Delta\mu_0 H$ = 5 T) were observed in Ce(Fe$_{0.96}$Ru$_{0.04}$)$_2$ [50]. Further substitution of Fe by Ru caused a strong deterioration of the magnetocaloric effect and $\Delta S$ decreased for Ce(Fe$_{0.90}$Ru$_{0.10}$)$_2$ to about -4 Jkg$^{-1}$K$^{-1}$ (AFM-FM) and 2 Jkg$^{-1}$K$^{-1}$ (FM-PM) [50]. In general, entropy changes for Ce(Fe$_{1-x}$M$_x$)$_2$, where M is metal, for $\Delta\mu_0 H$ = 5 T, in most cases ranged from 3 Jkg$^{-1}$K$^{-1}$ to 5 Jkg$^{-1}$K$^{-1}$ [49-52]. In conclusion, the presence of disorder in the measured samples caused partial suppression of structural transition (AFM-FM) and broadening of the FM-PM transformation range, which finally led to a diminution in isothermal entropy changes.

**Conclusions**

The Ce(Fe$_{0.9}$Co$_{0.1}$)$_2$ compound synthesized in a single crystalline form or even as a polycrystal by arc-melting/annealing procedure may be treated as the quasi-equilibrium counterpart of the samples analyzed in this paper. In such a homogenized state, Ce(Fe$_{0.9}$Co$_{0.1}$)$_2$ is known to undergo the AFM-FM magnetostructural transition with an antiferromagnetic state stable at low temperatures [11]. The stabilization of the FM phase at low temperatures at the expense of the AFM one is observed here for an as-quenched sample already. The quenched-in topological disorder accompanying rapid quenching is responsible for such a strong effect. As expected, severe plastic deformation further stabilizes the FM phase at low temperatures. The analyzed compound is on the verge of being ferromagnetic and even small instabilities (*e.g.* structural disorder, or applied magnetic field) force the emergence of the FM phase. Similar behavior was reported by us before for the YCo$_2$ Laves compound [21], where a ferromagnetic ordering was induced in an exchange-enhanced Pauli paramagnet. Other possible origins of such strong emergence of ferromagnetism, such as the presence of various structural units (related to small distortions of MgCu$_2$-type structure), were analyzed based on the *ab initio* calculations also. The results indicate that the AFM state is stable irrespective of the structural distortion of the initial MgCu$_2$-type structure, neglecting such origin as being responsible for the emergence of the FM phase in our case. Regarding the magnetocaloric effect, the structural disorder reduced the volume fraction transformed from a ferromagnetic to an antiferromagnetic state. Therefore, a strong reduction of *ΔS* (compared to the as-quenched sample) was observed. Moreover, this reduction is not compensated by a large broadening of the $\Delta S(T)$ peaks, so the structural disorder diminishes both the magnetocaloric effects associated with the AFM-FM and FM-PM transformations.

**Acknowledgments**

This work received financial support from the National Science Center Poland under grant DEC-2021/41/B/ST5/02894 (OPUS 21). The authors thank Julia Ivanisenko for making the high-pressure torsion experiments possible and for revision. The authors would like to thank Mirosław Werwiński for helpful discussions and critical reading.

**Conflict of interest**

The authors declare that they have no conflict of interest.

**Appendix**

Table 1. Lattice parameters of stable AFM variants of supercells of $Ce(Fe_{0.875}Co_{0.125})_2$.

| Symmetry Group | a [Å] ±0.001 | b [Å] ±0.001 | c [Å] ±0.001 | α ±0.01 | β ±0.01 | γ ±0.01 |
|---|---|---|---|---|---|---|
| *Cmm*2 | 7.185 | 7.195 | 7.195 | 89.72 | 90.00 | 90.00 |
| *C*222$_1$ | 7.190 | 7.190 | 7.190 | 90.00 | 90.00 | 90.06 |
| *P*2/*m* | 7.192 | 7.192 | 7.190 | 89.91 | 90.09 | 90.09 |
| *Cmm*2 | 7.195 | 7.195 | 7.185 | 90.00 | 90.00 | 90.29 |
| C222$_1$ | 7.190 | 7.189 | 7.190 | 90.00 | 89.94 | 90.00 |
| *Cmm*2 | 7.195 | 7.186 | 7.195 | 90.00 | 89.73 | 90.00 |